\documentclass[11pt]{article}

\usepackage{amsmath}
\usepackage{amssymb}
\usepackage{bm}
\usepackage[dvipsnames]{xcolor}
\usepackage[margin=1in]{geometry}
\usepackage{graphicx}
\usepackage[
urlcolor=blue
]{hyperref}
\usepackage{microtype}
\usepackage{natbib}

\usepackage{todonotes}

\usepackage{changes} 
\definechangesauthor[color=cyan]{JJ}
\definechangesauthor[color=blue]{RM}
\definechangesauthor[color=red]{MAP}

\hypersetup{
colorlinks=true,
citecolor=blue,
} 

\title{Neuromorphic architectures as numerical solvers\\for computational neuroscience}
\author{Jakob Jordan$^{1,2}$\footnote{Corresponding author:~\href{mailto:jakob.jordan@yale.edu}{jakob.jordan@yale.edu}}, Ole Richter$^{3,1}$, Congyang Li$^{1}$, Mihai A.~Petrovici$^{2}$\footnote{Joint senior authorship.}, Rajit Manohar$^{1\dagger}$\\[1em]
  \small $^1$Department of Electrical and Computer Engineering, Yale University, New Haven, CT, USA\\[1ex]
  \small $^2$Department of Physiology, University of Bern, Bern, Switzerland\\[1ex]
  \small $^3$Department of Applied Mathematics and Computer Science,\\ \small Technical University of Denmark, Kgs.~Lyngby, Denmark\\
}
\date{\today}

\begin{document}

\maketitle

\begin{abstract}
Neuromorphic computing is closely associated with spiking neuronal networks.
However, an alternative class of so-called ``rate-based'' models arising from computational neuroscience and machine learning forgoes spiking interactions and instead relies on continuous coupling between neurons.
Existing neuromorphic implementations designed around spike-based interactions are not well-suited for emulating such models.
Here we view the distributed simulation of these models as message-passing algorithms on parallel hardware.
Leveraging prior art in numerical algorithms and distributed simulation, we outline steps that enable the design of efficient digital neuromorphic accelerators for non-spiking neuronal models. 
In particular, we show that multi-bit packets, rather than spikes, are the most efficient communication strategy in packet-switched networks and that compared to basic numerical integration methods, higher-order differential equation solvers decrease both computation and communication costs while achieving lower numerical error, but that these benefits are ultimately limited by arithmetic precision.
Using our proposed design principles, we convert an existing neuromorphic architecture into a distributed numerical solver -- a spikeless neuromorphic system -- for continuously-coupled neuronal models.
We thereby demonstrate that our theoretical considerations indeed translate into practical advantages, namely reduced energy consumption and delay.
\end{abstract}

\section{Introduction}

Neuromorphic computing revolves around the efficient implementation of bio-inspired computation, with the brain as a prime example.
The field originally focused on exploiting analog features of electronic circuits for efficient implementations of computational primitives \citep{mead1990neuromorphic,douglas1995neuromorphic}, with modern systems comprising millions of analog circuits \citep{schemmel2010wafer,benjamin2014neurogrid,neckar2018braindrop}.
Over the last decade, researchers have also developed fully digital neuromorphic systems \citep{furber2014spinnaker,merolla2014million,davies2018loihi}, motivated by better scaling with improving manufacturing technology, as well as increased robustness to process variations and noise \citep{furber2024digital}.

\begin{figure}[t]
\centering
\includegraphics[width=1.0\textwidth]{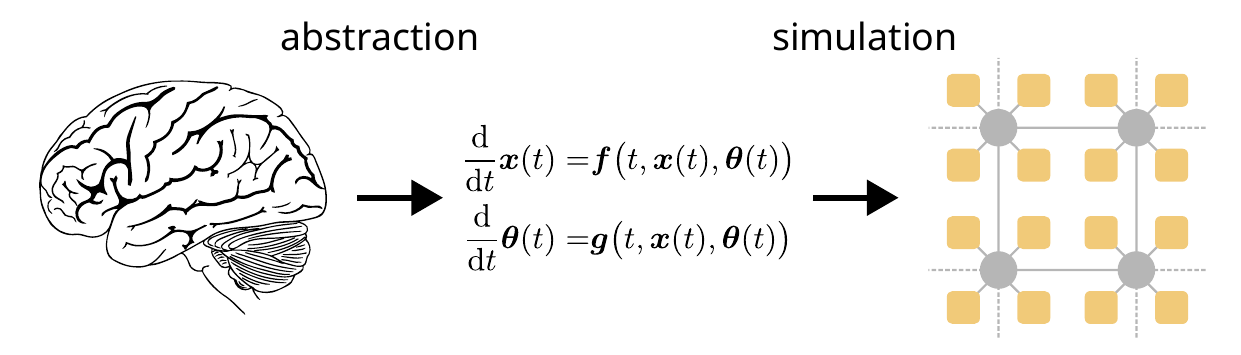}
\caption{{\bf Neuromorphic systems are distributed ordinary differential equation solvers.}
  To simulate brain dynamics and function, the vast majority of biological complexity is abstracted away, resulting in sparse systems of ordinary differential equations with few equations per neuron and synapse.
  Here $\bm{x}$ denotes neuronal states, for example membrane potentials, while $\bm{\theta}$ denotes parameters, for example synaptic weights.
  Neuromorphic systems are used to accelerate the simulation of their dynamics and/or reduce the energy expenditure of doing so.
  The right panel shows a sketch of a typical neuromorphic many-core architecture, with cores in orange and routers and links in gray.
}\label{fig:brain-to-sim}
\end{figure}

Since the biological wetware is too complex to simulate directly \citep{peters1991fine,squire2003memory}, researchers build abstractions, i.e., phenomenological descriptions that ignore physiological details currently believed to be unimportant.
Focusing on computation, the community has largely agreed that the most interesting dynamics can be captured at the neuronal and synaptic modeling level \citep{dayan2005theoretical,gerstner2014neuronal}.
Taking inspiration from other natural sciences, these descriptions typically take the form of ordinary differential equations (ODEs).
The resulting non-linear dynamical systems are usually not analytically solvable, and researchers use numerical simulations to study their behavior (Fig.~\ref{fig:brain-to-sim}).
In this context, modern neuromorphic systems can be seen as distributed differential equation solvers, specifically designed to accelerate the simulation of large-scale neuroscientific models.
However, all current neuromorphic systems are designed for the simulation of spiking neuronal networks.

In spiking neuronal networks, the spiking non-linearity and subsequent reset introduces significant temporal sparsity \citep{maass1997networks}.
By leveraging (digital) multiplexed communication \citep{liu2014event,moradi2018impact} this temporal sparsity can be directly exploited and allows the design of efficient architectures despite the high-fan out of biological neurons \citep{stepanyants2009fractions}.
The resulting systems are consequently an excellent fit for algorithms defined on sparse binary events, such as barn-owl-inspired directional hearing \citep{sullivan1986neural} building on the Jeffress model \citep{jeffress1948place}, spike-wave processing inspired by weakly electric fish \citep{kawasaki1988temporal,engelmann2016modeling}, temporal implementations \citep{ponulak2013rapid,aimone2019dynamic,davies2021advancing,li2025deterministic} of graph search \citep{lavalle2006planning}, stochastic approaches to quadratic unconstrained binary optimization \citep{alom2017quadratic,mniszewski2019graph}, sampling-based probabilistic inference \citep{petrovici2016stochastic,kungl2019accelerated}, and solving classification problems using temporal codes \citep{mostafa2017supervised,goltz2021fast}.
As a result, executing them on neuromorphic systems yields significant efficiency gains compared to general-purpose machines \citep{davies2021advancing}.
However, the specific implementation constraints of spiking neuromorphic systems are challenging to design for, resulting in a lack of performant and scalable algorithms \citep{davies2021advancing,schuman2022opportunities,kudithipudi2025neuromorphic,muir2025road}.

This view on neuronal computations and the resulting implementations rely on a specific abstraction:~spikes are represented as binary events at single moments in time with no spatial or temporal extent.
However, many approaches in computational neuroscience avoid this abstraction and instead use systems of non-linear differential equations with continuous coupling.
This includes seminal spiking neuron models \citep{hodgkin1949effect,fitzhugh1961impulses,nagumo1962active}, but also modern approaches focused on capturing compound activity \citep{xie2003equivalence,lee2015difference,scellier2017equilibrium,sacramento2018dendritic,song2020can,meulemans2021credit,haider2021latent,laborieux2022holomorphic,fayyazi2024prospective,ellenberger2025backpropagation,zucchet2025teaching}, also explored in machine learning \citep{chen2018neural,voelker2019legendre,gu2020hippo,hasani2021liquid,zucchet2023online,gu2024mamba}.
The notable success of the latter model class in solving common signal processing tasks may motivate their emulation in existing spiking hardware.

\begin{figure}[t]
\centering
\includegraphics[width=0.99\textwidth]{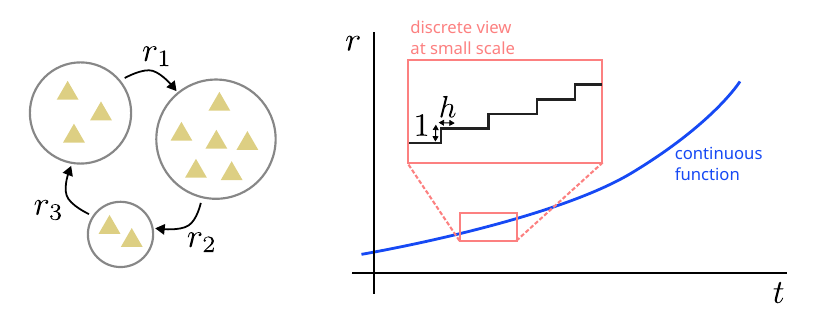}
\caption{{\bf The interactions of continuous models may be reduced to binary events for small time steps.}
    Neuronal models, for example neuronal population models are formulated as systems of ODEs with continuous interactions.
    Decreasing the observation time scale in digital implementations allows the use of binary events for communication of rates.
    Here $r$ denotes the ``firing rate'' of a population, $t$ (simulated) time, and $h$ the simulation time step.
}\label{fig:cont-to-spikes}
\end{figure}

An obvious first step would be to simply replace the spiking nonlinearity with a continuous one and, to retain the communication paradigm via binary events, implement a variant of delta dataflow \citep{manohar2004dataflow}, also known as send-on-delta, or sigma-delta coding \citep{oconnor2016sigma}, i.e., use events to signal changes.
Naturally this assumes a certain quantization of values in line with the assumption of digital communication.
This communication style is indeed a good choice if most of the time nothing happens, i.e., when the speed of the system's components (the ADC rate in mixed-signal systems; the inverse time-step size in digital systems) is much faster than the relevant time scales of the model.
The combination of such time-scale differences with send-on-delta coding results in temporally sparse communication (Fig.~\ref{fig:cont-to-spikes}).
However, to preserve high accuracy, the time steps would need to be exceedingly small, leading to long run times.
One way to accelerate the time to solution is by increasing the integration step size while remaining within error bounds defined by the application requirements.
Yet, in this regime, it is not sufficient to communicate binary changes any more -- since multiple quantization levels may be crossed in a single step, we have to communicate values with higher resolution than simply one bit.

In the following we develop design principles for neuromorphic systems specifically tailored to accelerating continuously-coupled models.
First, we show that multi-bit packets are the most efficient communication strategy in packet-switched networks.
Second, we demonstrate that higher-order ODE solvers decrease both computation and communication costs while achieving lower numerical error, but that these benefits are ultimately limited by arithmetic precision.
Finally, using our proposed design principles, we convert an existing neuromorphic architecture into a distributed ODE solver for continuously-coupled models and obtain pre-layout energy and delay estimates.

\section{Spikes are (only) a good choice for communicating binary events}

In the following, we start from a modern digital neuromorphic multi-core architecture with a network on chip and time synchronization that ensures deterministic execution \citep{merolla2014million,davies2018loihi,li2025deterministic}.
The interaction between different variables in a continuously-coupled system of ODEs requires the communication of multi-bit values.
We can split the proposed solutions to this problem into two classes, depending on how they scale with respect to resources.
Since we are interested in communicating a value from one neuron to another, we only consider the scaling with respect to the number of time bins and ignore codes which use ``spatial'' encodings across multiple neurons, including rank-order coding \citep{thorpe1998rank} and group neurons \citep{lv2024optimal} (see Appendix Sec.~\ref{sec:methods-codes} for additional details).

\begin{figure}[t]
\centering
\includegraphics[width=1.0\textwidth]{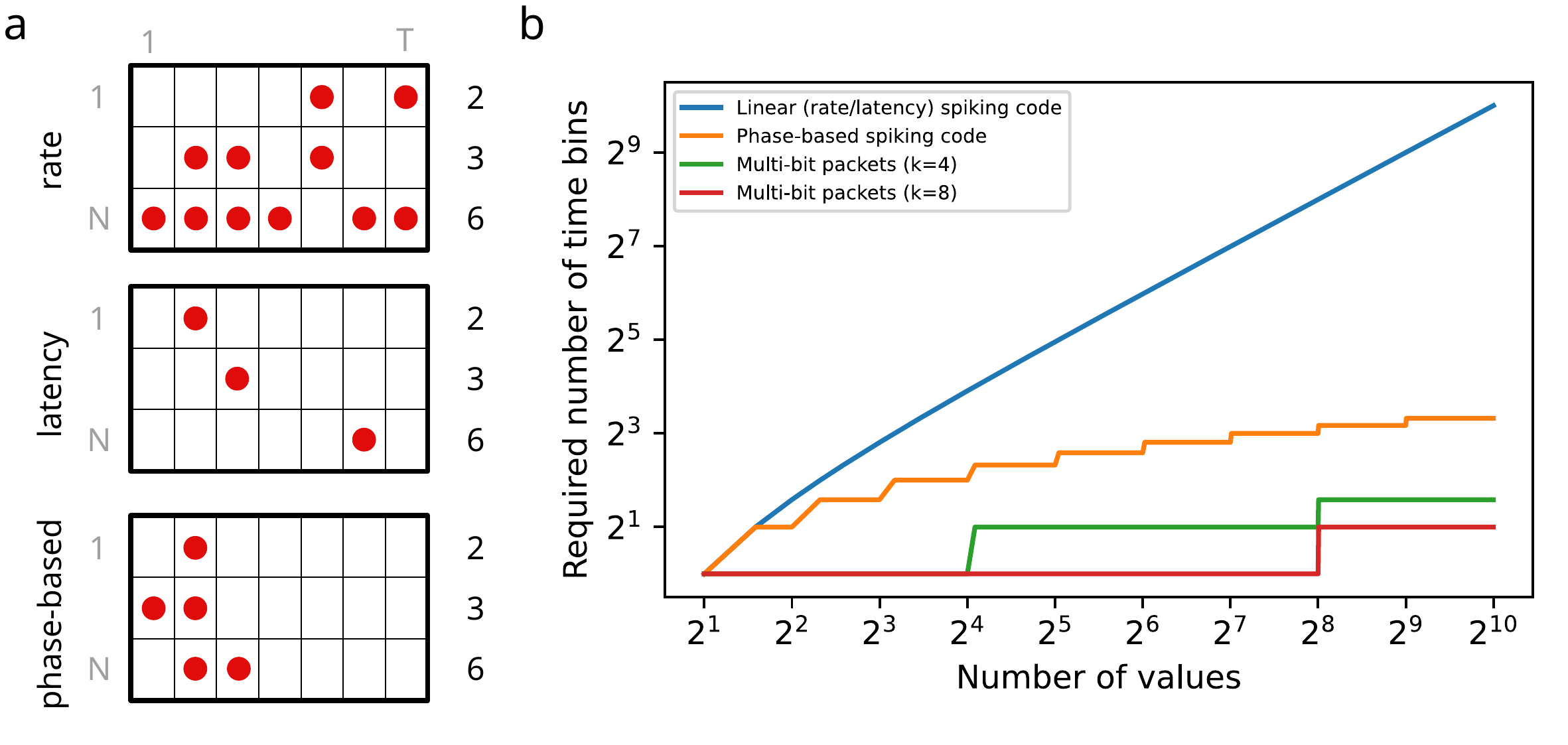}
\caption{{\bf Multi-bit packets are most efficient to communicate non-binary values.}
{\bf (a)} Illustration of different spiking codes.
$N$ indicates the number of neurons, $T$ the number of time bins.
Here $N=3, T=7$.
Numbers on the right denote the encoded value.
{\bf (b)} Required number of time bins as a function of number of values for a population size of $N=1$ for different coding schemes.
}\label{fig:costs-of-communication}
\end{figure}

Due to time synchronization, cores agree on a common temporal frame of reference and the presence or absence of a spike in a time bin allows communicating $1$b of information.
{\it Linear\/} spiking codes can thus encode $T+1$ values in $T$ time bins and require resources proportional to the number of different values (Fig.~\ref{fig:costs-of-communication}).
Examples of this approach include rate codes which transmit a number of spikes proportional to the encoded value \citep{rueckauer2016theory,tang2017sparse,han2020rmp,bu2023optimal,xu2026neuromorphic}, time-to-first-spike codes which use the latency from some global reference time to encode values \citep{rueckauer2018conversion,park2020t2fsnn,frady2020neuromorphic,stanojevic2024high}, and codes which utilize interspike intervals \citep{han2020deep}.
Attempts to speed up inference in networks using such linear codes \citep[e.g.,][]{bu2023optimal} can not remove this fundamental scaling property and any improvements can be traced back to using stronger quantization and quantization-aware training \citep{voelker2020spike}.
Note that we do not consider stochastic codes, such as Poisson rate coding, as they perform significantly worse than linear codes if not operated in an extremely quantized setting \citep{thorpe1998rank,manohar2015comparing}.

{\it Combinatorial\/} spiking codes, in contrast, scale more favorably in terms of resources.
For example, phase-based codes \citep[e.g.][]{kim2018deep,zhang2020efficient} implement bit-serial communication of binary representations, and can therefore encode $2^T$ values in $T$ time bins, achieving logarithmic scaling (Fig.~\ref{fig:costs-of-communication}).

Can we further improve over spiking combinatorial codes in engineered systems?
In packet-switched communication strategies, such as the commonly used address-event representation \citep{mahowald1992vlsi}, packets contain typically $16$b-$32$b routing information (``header'') \citep{furber2013overview,thakur2018large}.
In these systems, adding a small ``payload'' of $B$ bits causes negligible additional costs.
This thus allows us to communicate $2^{BT}$ values in $T$ time bins.
Compared to the phase-based code, this naturally results in a decrease in required resources that is inversely proportional to the payload size (Fig.~\ref{fig:costs-of-communication}).
These theoretical benefits can directly translate into practical advantages  --  a recent study found a five-fold decrease in synaptic operations per second using multi-bit packets \citep{khacef2025privacy}.

So far we have focused on the number of time bins required, i.e., the delay and {\it static\/} energy costs associated with transmitting a multi-bit value.
Considering {\it dynamic\/} energy costs, rate-based codes remain the worst choice.
First, the synaptic operation, i.e., multiplying the value with a synaptic weight, becomes more costly as it involves repeated additions, instead of a single efficient multiplication \citep[see also][]{ercegovac2004digital,davies2021advancing}.
Second, rate-based codes require the production and routing of the multiple events encoding a single value \citep{mostafa2017fast}.
Latency codes, in contrast, are maximally efficient in terms of dynamic energy, as they require at most a single event independent of the value being communicated.
Finally, multi-bit packets remain an improvement over phase-based codes since it is generally cheaper to route multiple bits jointly rather than individually due to reduced control overhead.
Since the dynamic energy costs of phase-based codes/multi-bit packets scales logarithmically in the number of values that need to be communicated, while the delay and static energy costs for latency codes exhibit linear scaling, the former remains likely the best choice for applications working outside of severely quantized regimes.

\section{Higher-order solvers increase temporal sparsity}\label{sec:higher-order}

The previous section was concerned with {\it how\/} to communicate multi-bit values between neurons; additionally, we need to decide {\it how often\/} to communicate.
In spiking systems this question is easy to answer:~the necessary communication frequency is determined by the frequency of spikes.
While continuous models ostensibly require continuous communication, numerical methods can create approximate solutions with known error bounds on a finite time grid.
Here we view neuromorphic systems as implementations of distributed differential equation solvers.
In particular, current digital spike-based neuromorphic hardware corresponds to a forward Euler solver \citep{hairer1993solving} implemented with fixed point arithmetic \citep{seo201145nm,cassidy2013cognitive,ma2017darwin,frenkel2019morphic,orchard2021efficient,richter2023speck}.
This, however, is not necessarily the best choice.
It is well-known in the literature of numerical methods for ODEs that more sophisticated algorithms can achieve higher accuracy if the number of computations performed is held constant.
Conversely, and important for the design of efficient systems, they can achieve equivalent accuracy while doing fewer computations.
Yet, these methods have so far received little attention in the design of neuromorphic systems.

Different solvers are classified by their {\it order}.
The order describes how the global integration error scales with the step size $h$:~a solver of order $p$ has global error $\epsilon \sim \mathcal{O}(h^p)$ \citep{hairer1993solving}.
For example, given a differential equation $\dot x = f(t,x(t))$, the forward Euler method with order $p=1$ advances time by $h$ using the update rules
\begin{align}
\notag
k_0 =& f(t, x(t)) \\
x(t+h) =& x(t) + h k_0 \;.
\end{align}
On the other hand, Ralston's method ($p=2$) uses the following update rules \citep{ralston1962runge}
\begin{align}
\notag
k_0 =& f(t, x(t)) \\
\notag
k_1 =& f(t + \frac{2}{3}h, x(t) + \frac{2}{3}h k_0) \\
\label{eq:ralston}
x(t+h) =& x(t) + h \left( \frac{1}{4} k_0 + \frac{3}{4} k_1 \right) \;.
\end{align}
Notice that to achieve order $p=2$, two evaluations of $f$, represented by $k_i$, are necessary: one at time $t$, and the other at time $t+2/3 h$.
In general, such ``Runge-Kutta'' methods require additional function evaluations at intermediate time points to achieve higher order.
These intermediate evaluations are typically referred to as ``stages'' of the solver, and the methods consequently also as ``multi-stage'' methods.

To compare different solvers of different order, rather than fixing a step size, we fix the tolerated error $\epsilon$.
For a given tolerated error and solver order, the step size scales as
\begin{align}
h(\epsilon, p) \sim \epsilon^{\frac{1}{p}} \;,
\end{align}
i.e., higher-order solvers can take larger time steps than lower-order solvers.
Under the assumption that the right-hand side of the differential equation ($f$) is significantly more expensive to compute than other operations of the solver, we can express the scaling of the number $n_f$ of function evaluations of $f$ as a function of the solver order $p$ and the tolerated error $\epsilon$ as (Appendix Sec.~\ref{sec:methods-nfevs})
\begin{align}
\label{eq:nfev-scaling}
n_f(\epsilon, p) \sim p T \epsilon^{-\frac{1}{p}} \;,
\end{align}
where $T$ is the integration time.
This function has a single minimum at $p = -\log(\epsilon)$ (Appendix Sec.~\ref{sec:methods-optimal-order}).
For example, if $\epsilon = 10^{-4}$, a 9th-order solver would minimize the number of function evaluations.
Consequently, for $\epsilon = 10^{-4}$ and $p < 9$, increasing the order of the solver decreases the total number of function evaluations and therefore computational costs.

\begin{figure}[t]
\centering
\includegraphics[width=0.98\textwidth]{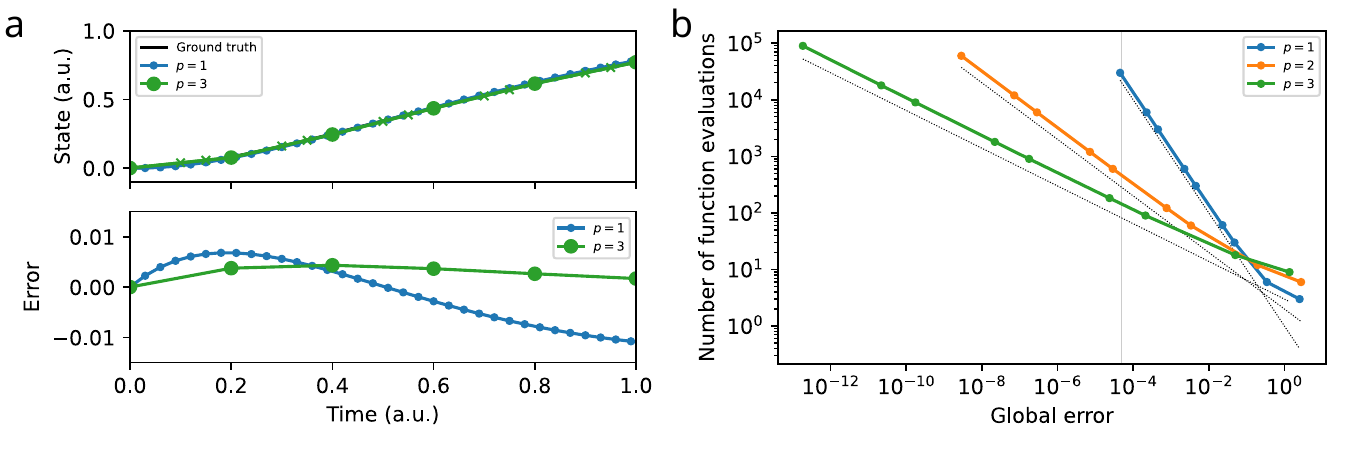}
\caption{{\bf Higher-order solvers require fewer function evaluations at fixed error.}
{\bf(a)} Trajectory of neuron (top) and error with respect to ground truth (bottom) in toy network.
Ground truth (black), forward Euler ($p=1$, blue), Ralston ($p=3$, green).
For visual clarity we do not include a second-order solver here.
Full circles are states at integer time steps (integer multiples of $h$), small crosses indicate intermediate evaluations.
Here $\tau = 0.5$, $h_{p=1} = 0.03$, $h_{p=3} = 0.2$.
{\bf (b)} Precision-work diagram.
Note that here ``precision'' refers to the accuracy of the integration method, not accuracy of the arithmetic type.
Dotted lines correspond to Eqn.~\ref{eq:nfev-scaling} for different solver orders.}\label{fig:benefits-of-higher-order}
\end{figure}

We investigate this scaling in a simple toy ``network'' consisting of two consecutive neurons modeled as low-pass filters without nonlinearities which can be solved analytically (Appendix Sec.~\ref{sec:methods-toy-model}).
For solvers of order one to three (Appendix Sec.~\ref{sec:methods-solvers}) the results match our theoretical predictions (Fig.~\ref{fig:benefits-of-higher-order}).
For a tolerated error of $5 \cdot 10^{-5}$ with respect to the analytical solution a third-order solver requires more than two orders of magnitude fewer function evaluations than a first-order solver.
Since each function evaluation requires information to be exchanged between coupled neurons, this simultaneously reduces the frequency of communication.

To put this into perspective, we can interpret the parameter $\tau$ in our example as a membrane time constant and consider the ``event frequency'' of different methods for matched error.
For $\tau=0.5$ms, a numerical error of approximately $5 \cdot 10^{-5}$ requires a time step of $h=0.0001$ms with a first-order method.
Since each state needs to be communicated, this results in a daunting event frequency ($\frac{1}{h}$) of $10$MHz assuming real-time operation.
A third-order method can achieve comparable numerical accuracy with a significantly larger time step ($h=0.05$ms).
Taking into account the necessary communication of intermediate stages, this results in an event frequency of $60$kHz -- a reduction of more than a factor of $160$.

Note that the above results do not account for the increased hardware cost for higher-order solvers, such as the additional memory required to store intermediate evaluations of the right-hand side ($f$) and the associated increased costs of loading neuronal state into registers.
We will consider these costs in Sec.~\ref{sec:neuroscale}.

\section{Higher-order solvers decrease required computations while minimizing errors in finite precision}\label{sec:lower-minimal-error}

Our previous results ignore the arithmetic error arising from finite bit widths in digital representations of continuous values.
In finite precision, the update of a state variable introduces an arithmetic error, for example in forward Euler arising from the addition of the scaled right hand side of the differential equation to the current state
\begin{align}
x(t + h) = x(t) + h f(t, x(t)) \;.
\end{align}
Over the course of a simulation of length $T$ this operation is performed $\frac{T}{h}$ times, and states thus accumulate a global rounding error of $\mathcal{O}(1/h)$, which increases with decreasing step size \citep[see also][]{kahan2013floating,dawson2018reliable}.
This scaling is opposite to the scaling of the integration error, which {\it decreases\/} with decreasing step size.
Consequently, these two contributions balance at some value, i.e., for a specific order, an optimal step size exists that minimizes the sum of both errors.

We assume that these errors combine linearly with some scaling factors $a$ and $b$:
\begin{align}
\label{eq:error-func}
\epsilon(h,p) =& a \epsilon_\text{order}(h,p) + b \epsilon_\text{precision}(h) \\
=& a T h^p + \frac{b T}{h} \;.
\end{align}
The minimum of this function with respect to $h$ is at (Appendix Sec.~\ref{sec:methods:optimal-step-size}, Appendix Fig.~\ref{fig:methods-optimal-step-size-error})
\begin{align}
h_\text{min}(p) = \left(\frac{c}{p}\right)^\frac{1}{p+1}\;,
\end{align}
where we introduced $c := \frac{b}{a}$.
Furthermore, $h_\text{min}(p)$ is a monotonically increasing function for (Appendix Sec.~\ref{sec:methods-larger-optimal-step-size})
\begin{align}
\label{eq:c-bound}
c < pe^{-\frac{1}{p} - 1} \;.
\end{align}
Since the right hand side of Eqn.~\ref{eq:c-bound} is a monotonically increasing function of $p\; \forall p \ge 1$, we plug in $p=1$ and conclude that for
\begin{align}
c < e^{-2} \approx 0.135 \;,
\end{align}
increasing the order of the integration method increases the optimal step size.
In other words, in this regime the required amount of computation that balances the contributions of rounding and integration error is lower for higher-order methods. 
We can furthermore show that the error at the optimal step size is lower for high-order methods (Appendix Sec.~\ref{sec:methods-smaller-error}).

\begin{figure}[t]
\centering
\includegraphics[width=0.99\textwidth]{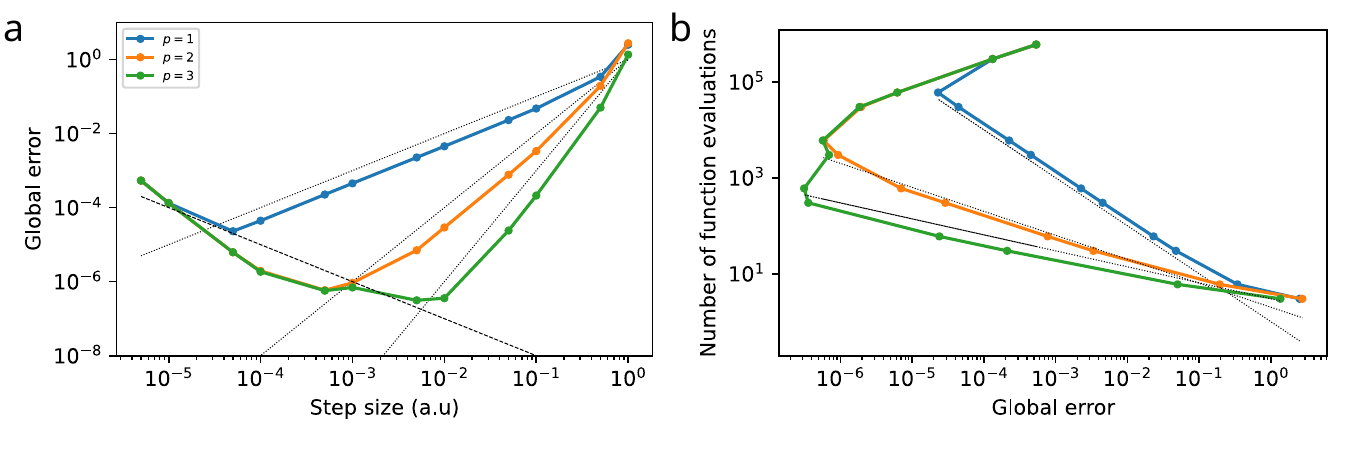}
\caption{{\bf Higher-order solvers achieve lower minimal error at fixed precision.}
{\bf (a)} Global error as function of step size.
Dotted lines indicate expected error scaling as a function of step size from solver order.
The dashed line indicates expected error scaling from arithmetic type (single-precision floating point, FP32).
{\bf (b)} Number of function evaluations as function of global error (compare Fig.~\ref{fig:benefits-of-higher-order}) for finite arithmetic precision.
Dotted lines correspond to Eqn.~\ref{eq:nfev-scaling} for different solver orders.
}\label{fig:precision}
\end{figure}

While $a$ depends on the properties of the system of ODEs and solver order (e.g., the second derivative for forward Euler), $b$ is determined by the chosen precision and expected value range.
Roughly, if state variables are of order $\mathcal{O}(1)$, it corresponds to the so called ``machine epsilon'', i.e., the next representable floating point value after $1$.
We reduce the precision in our toy simulation from Sec.~\ref{sec:higher-order} to single-precision floating point.
As expected, we observe two qualitatively different regimes:~for small step sizes errors are dominated by rounding error, and for large step sizes by integration error (Fig.~\ref{fig:precision}).
Finally, we estimate the coefficients $a, b$ by fitting Eqn.~\ref{eq:error-func} to numerical results.
We obtain $a \approx 1, b \approx 10^{-9}$, i.e., $c\ll 0.135$, suggesting that, in the absence of severe quantization, higher-order methods decrease the amount of required computations while simultaneously minimizing numerical errors.

\section{NeuroScale case study}\label{sec:neuroscale}

\begin{figure}[htp]
\centering
\includegraphics[width=0.85\textwidth]{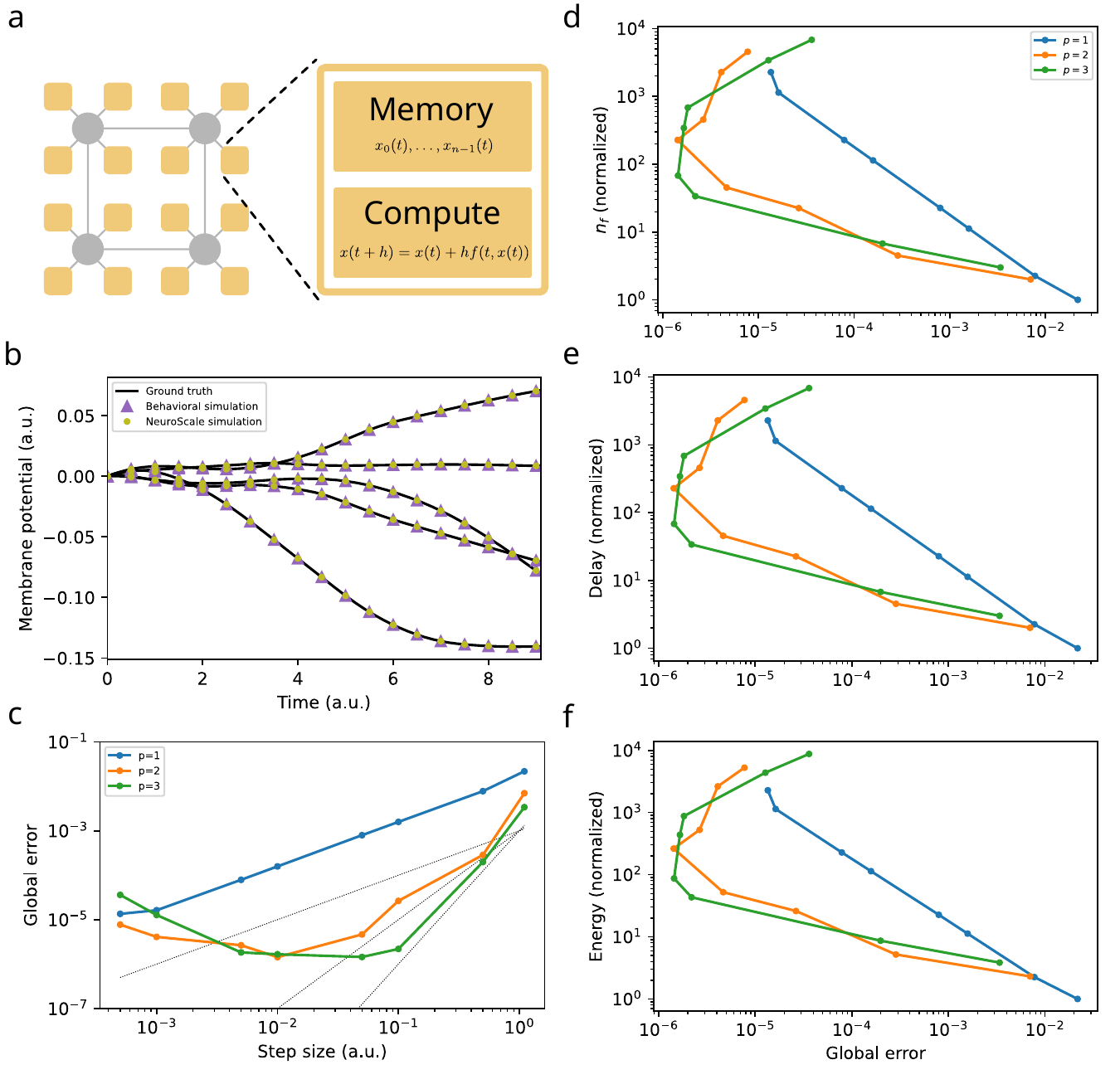}
\caption{{\bf Back-annotated simulations of a neuromorphic system confirm practical advantages of higher-order methods.}
{\bf (a)} Chip-level architecture of the simulated hardware implementation.
{\bf (b)} Trajectories of five example neurons of one realized random network (Appendix Sec.~\ref{sec:methods-neuroscale-model}) with $h=0.01$.
Black:~ground truth (SciPy's DOP853 solver), triangle and circle symbols correspond to behavioral and NeuroScale simulation, respectively (RK3 solver; Appendix Sec.~\ref{sec:methods-solvers}).
Only every fiftieth state is shown.
{\bf (c)} Scaling of global error with step size. Dotted lines indicate scaling expected from solver order.
{\bf (d-f)} Precision-work diagrams.
In each panel, we normalize all curves to the respective value of RK1 at the largest simulated step size.
}\label{fig:neuroscale}
\end{figure}

To illustrate the practical consequences of our theoretical results, we implement multi-bit packets and propagation of neuronal dynamics by higher-order solvers in a recent digital neuromorphic architecture \citep[NeuroScale;][]{li2025deterministic,li2026full} and simulate continuously-coupled models.
NeuroScale consists of a two-dimensional array of cores connected by a packet-switched mesh network on chip with deterministic dimensional order routing.
Each core has local memory to store neuron states and synaptic weights, along with circuits for the corresponding computations (Fig.~\ref{fig:neuroscale}a).
Similar to other digital neuromorphic systems, synaptic accumulation is event-driven, while neuronal states, i.e., membrane potentials, are propagated on a regular time grid using a forward-Euler-like solver.
Determinism is ensured by a pairwise time-step-level synchronization between cores.
The original NeuroScale implementation uses the CHP (Communicating Hardware Processes) language \citep{martin1993synthesis}, a synthesizable high-level language part of the ACT (Asynchronous Circuit Toolkit).
We adapt the original implementation as described below and obtain delay and energy estimates via back-annotated simulations of the synthesized design \citep[Appendix Sec.~\ref{sec:methods-neuroscale-simulations};][]{srinivasan2025maelstrom,srinivasan2026improved}.

Due to the increased complexity of both neuronal dynamics and synaptic interactions as well as increased memory footprint of higher-order solvers, a single core of our modified design can accommodate fewer neurons than the original implementation.
Assuming $128$ neurons per core, the simulated system consisting of $16$ cores could simulate the dynamics of up to $2048$ neurons.
Reaching the number of neurons required by large-scale cortical models can be achieved by scaling the system in two directions.
Besides scaling the number of cores per chip, off-chip interfaces can extend the network on chip hierarchically, similar to the approach followed by Loihi \citep{davies2021advancing}.

Each local state may be influenced by states stored remotely.
Computing the right-hand side $f_i(t, \bm{x})$ of each neuron's differential equation thus requires communication between cores.
For example, for a first-order (forward Euler) solver, at each time step, neurons first need to communicate their state to all other cores on which they have targets.
We can then compute the first (and only) stage for each neuron and correspondingly update its local state, after which we proceed to the next time step.
Higher-order solvers require intermediate evaluations ($k_i$), which similarly require communication before each stage can be computed. 
To implement higher-order solvers with multiple stages without altering the existing communication protocol, we reinterpret the (integer) steps of the original implementation as follows:~for a solver of order $p$, neuronal states at steps $i\!\!\mod p = 0$
correspond to neuronal states at time step $i / p$, while other steps are used for the computation of intermediate stages and the corresponding intermediate states.
For example, for Ralston's method (a second order solver, see Eqn.~\ref{eq:ralston}), steps $i=0,2,4$ correspond to the neuronal states at times $t=0, 1, 2$ and are used to compute the first stage ($k_0$), while steps $i=1,3,5$ compute the second stage ($k_1$) and propagate neuron states.
Since later stages may depend on linear combinations of previous stages, we linearly increase the memory for neuronal states with the number of stages.

We use mixed-precision fixed-point arithmetic with a higher precision for neuronal states (Q$8.24$), and lower precision for communication and computation (Q$4.18$), with ``round to nearest, ties to even''.
We correspondingly adapt the packet format to include additional payload bits.
We compare Runge-Kutta methods of order one (forward Euler; RK1), two (Ralston; RK2), and three (Ralston; RK3), simulating a recurrent network of leaky-integrator neurons with ReLU nonlinearity (Appendix Sec.~\ref{sec:methods-neuroscale-model}) and a random sparse connectivity matrix using Kaiming initialization \citep{he2016deep}.
Ground truth trajectories are obtained from SciPy's ``DOP853'' solver with $rtol=atol=10^{-12}$ \citep{virtanen2020scipy}.

The trajectories obtained from back-annotated simulations are bitwise identical to a custom behavioral simulator using standard sequential implementations of the respective solvers (Fig.~\ref{fig:neuroscale}b).
For the scaling of the error with time step size, we observe similar results as for the limited-precision toy example (Sec.~\ref{sec:lower-minimal-error}):~decreasing the step size decreases errors faster for higher-order methods, yet at some point the arithmetic errors due to finite precision start to dominate (Fig.~\ref{fig:neuroscale}c).
The (normalized) delay follows the same trend as the (normalized) number of function evaluations (Fig.~\ref{fig:neuroscale}d,e).
This may not be entirely surprising:~due to the sequential dependencies between states at different time points, a larger (smaller) number of function evaluations directly translates linearly into longer (shorter) use of the available hardware resources and parallelism across the system, i.e., propagating multiple states simultaneously for one step, does not change this scaling.
For (normalized) energy, the wider memories required by higher-order solvers result in an offset without changing the previously observed behavior (Fig.~\ref{fig:neuroscale}f).
Jointly these results demonstrate that our theoretical considerations directly translate to practical advantages.

\section{Conclusion}\label{sec:conclusion}

\paragraph{Summary}
We proposed to view neuromorphic systems as distributed ODE solvers.
Through this lens, we have rigorously studied the resulting trade-offs between complexity and efficiency and have translated our findings into specific design principles tailored to models with continuous interactions.
First, we have argued for multi-bit packets as the most efficient choice in packet-switched networks; indeed some recent neuromorphic systems are already evolving into that direction \citep{orchard2021efficient}.
Second, we have highlighted how well-known numerical methods can provide significant reductions to both computation and communication.
Finally, we have provided a proof-of-concept implementation and obtained empirical measurements that support our theoretical arguments.

\paragraph{Precision matters}

Higher-order methods are more efficient than lower-order methods only for sufficiently small error tolerance (Fig.~\ref{fig:benefits-of-higher-order}).
From this observation, the question arises whether small numerical errors are sufficiently necessary to justify the corresponding implementation costs.
To address this, let us distinguish two different use cases of neuromorphic hardware:~(i) simulation of phenomenological models and (ii) machine-learning-like applications.

In the first case, one aims to explore the properties of a mathematical model using simulations.
Small numerical errors ensure that the simulated trajectories accurately reflect the properties of the mathematical model as opposed to those of the specific numerical implementation.
In contrast, large numerical errors easily lead to a violation of basic properties of systems of ODEs such as uniqueness of trajectories \citep{dupont2019augmented,ott2020resnet}, or worse, can silently lead to a misinterpretation of results \citep{pauli2018reproducing}.

For the second case it might be tempting to argue that numerical errors are irrelevant as long as a task is successfully learned, i.e., some objective function maximized, and thus to conclude that simpler integration methods are sufficient or even superior.
This attitude neglects the practical benefits of being able to transfer expectations and insights between theory and implementation, in particular in order to be able to trace and separate both algorithm failures and implementation bugs throughout the development cycle.
Furthermore, small numerical errors increase the portability of parameters obtained from simulations involving learning.
As the performance of learning algorithms increases, so does the risk of overfitting model parameters to the specific (errors of the) implementation \citep{gusak2021meta}.
While small numerical errors can not prevent this effect, they can help mitigate it.
Finally, large numerical errors can be detrimental to the convergence of learning \citep{onken2020discretize,peng2023fp8,creswell2024understanding}.
We note that certain optimizations, such as the introduction of additional quantization introduced by delta-sigma coding/send-on-delta \citep{cheung1993sigma,oconnor2016sigma}, introduce additional numerical errors which need to be measured and controlled for.

More broadly speaking, when creating an artificial system, we have the opportunity to design it in a way that allows us to reliably reason about it.
One way to achieve this is through abstraction.
Small numerical errors isolate the modeling level from implementation details, allowing a reliable abstraction prior to manufacturing and deployment.

\paragraph{Consequences for the design of spiking neuromorphic hardware}
Here we focused on continuously-coupled systems of ODEs, and argued that higher-order solvers decrease the frequency of communication and thus reduce both delay and energy.
In contrast, in spiking neuronal networks the communication frequency is not determined by the integration step size, but rather by threshold crossings, i.e., it is a property of the model, not its numerical solution.
For models that can be solved in closed form \citep{rotter1999exact} one may even entirely forgo numerical integration and directly implement a discrete-event simulation \citep{thomas2000parallel,goltz2021fast,engelken2023sparseprop}.
For the integration of {\it nonlinear\/} subthreshold dynamics, spiking implementations would also enjoy the benefits of higher-order solvers to reduce the amount of necessary computation.
Yet, the larger step sizes associated with this reduction cause new challenges, requiring interpolation and root finding methods to avoid restricting spike times to coarse time grids \citep{hanuschkin2010general} and to avoid missing threshold crossing \citep{krishnan2018perfect}.
Determining whether the savings of higher-order solvers outweigh the costs imposed by these (necessary) complexities is essential for the construction of more efficient spiking neuromorphic hardware.

\paragraph{Relation to partial differential equation (PDE) accelerators}
The design principles for neuromorphic ODE solvers presented here are related to a parallel research theme exploring ASICs for the accelerated solution of PDEs.
Similar to neuromorphic architectures, an initial focus on mixed-signal implementations \citep{cowan2006vlsi,guo2016energy,natarajan2020analog,udayanga2020radio,liang2021offset} has gradually given way to fully digital architectures \citep{kung2017programmable,chen2020accelerator,huang2024intiacc,shamsi2024implementation,mu2024scalable,ju2024physics}.
These digital implementations aim to reduce time- and/or energy-to-solution by parallelizing work and reducing data movement using arrays of simple cores equipped with a suitable interconnect, in a similar spirit as the architectural family discussed here.
Yet, two crucial differences between PDE accelerators and neuromorphic architectures arise from the typical ``connectivity'' of the respective model classes.

First, PDE accelerators can exploit the regular structures arising from the spatial discretization of the underlying partial differential equation systems.
The typical method (method of lines via finite differences) leads to sparse, yet perfectly uniform connectivity resulting in predictable dataflow that can be directly optimized for.
This approach fits well to physics problems with local coupling and spatial dimensionality matched to the dimensionality of the interconnect.
However, the sparse and irregular connectivity of cortical models \citep{braitenberg2013cortex,senk2022connectivity} requires more flexible communication topologies that support in principle arbitrary communication patterns between cores.
A similar requirement arises in recent accelerators for so called ``Ising machines'', a physics-inspired approach for solving constraint satisfaction problems \citep{su2024flexspin}.

Second, the high fanin of cortical models \citep{schuz1989density,stepanyants2009fractions} with potentially plastic, weighted connections requires specific optimizations in neuromorphic systems \citep{moradi2018impact,li2025deterministic} absent from PDE accelerators.
For example, in NeuroScale \citep{li2025deterministic} a neuron can receive weighted input from up to $4096$ other neurons located on arbitrary cores in the system.
In stark contrast, a grid point in a typical core optimized for a two-dimensional PDE receives unweighted input from its four direct neighbors \citep[e.g.,][]{chen2020accelerator}.

\paragraph{Outlook}
The simulation of continuously-coupled models calls for new neuromorphic architectures that take their differences to spiking models into account.
The resulting systems will be well suited for temporal algorithms with continuous interactions and thereby complement spiking neuromorphic systems tailored to algorithms defined on sparse binary events.
Even after adopting the necessary changes discussed here, the resulting implementations will benefit from previous designs.
Their communication patterns – small packets requiring routing to a specific fraction of the whole system – are closely related to existing neuromorphic principles and in particular a good fit for asynchronous digital implementations.

Here we discussed {\it what\/} to communicate and {\it how frequently}, the next step is to consider {\it when\/} to communicate.
We assumed a fixed time step size for the whole system, requiring the worst choice across space and time.
Adaptive step sizes can avoid the corresponding unnecessarily high computation and communication load by propagating each neuron on its own optimal time grid.
Furthermore, they allow for online error control.
Adaptive time steps create new challenges as the states of inputs are not necessarily available at times a neuron is updated, requiring, for example, interpolation techniques.
Yet, the resulting implementations of continuous models will exhibit significant temporal sparsity, a property that is frequently -- but incorrectly -- uniquely assigned to spiking neuronal networks.

\paragraph{Acknowledgements}
We gratefully acknowledge funding from the European Union under grant agreement \#101147319 (EBRAINS 2.0), from the Swiss National Science Foundation (Grant Number 233569) and from the United States National Science Foundation (NSF Award Number 2449506). We owe a particular debt of gratitude for the ongoing support from the Manfred Stärk Foundation.
We thank Karthi Srinivasan for always-on support with circuit synthesis.

\paragraph{Data availability}
The underlying code and data for Fgs.~3-5 and Fig.~7 are available at \href{https://codeberg.org/jakobj/neuromorphic-architectures-numerical-solvers}{https://codeberg.org/jakobj/neuromorphic-architectures-numerical-solvers}. 
The underlying code and data for Fig.~6 is not publicly available but may be made available by the corresponding author to qualified researchers on reasonable request to the degree permissable by licensing agreements.

\clearpage

\small
\bibliographystyle{apalike}
\bibliography{bib}

\clearpage

\appendix

\section{Spiking codes}\label{sec:methods-codes}

The possible number of encoded values for different codes are
\begin{itemize}
\item rate/latency:~${(T+1)}^N$
\item rank-order:~$N!$ (assuming $T \geq$ N)
\item phase-based:~$2^{TN}$
\end{itemize}
where $T$ denotes number of time bins and $N$ number of neurons.

The least informative distribution over $N$ neurons and $T$ time bins is the uniform distribution with entropy
\begin{align}
\notag
S =& -\sum_i p_i \log_2 p_i \\
=& -\sum_i \frac{1}{2^{NT}} \log_2 \frac{1}{2^{NT}} \\
=& NT \;.
\end{align}
Comparing this to the code rate of phase-based codes ($NT$) suggests that only phase-based codes make optimal use of the available resources \citep{shannon1948mathematical}.
The lower code rate of other codes results from redundant representations (rate coding and rank order coding) and unused symbols (latency coding).

If we allow $B$ bits of payload we can encode $2^{BTN}$ different values.
For a single neuron, i.e., $N=1$, the required number of time bins to communicate $v$ different values are thus
\begin{itemize}
\item rate/latency:~$v-1$
\item phase-based:~$\log_2 v$
\item multi-bit:~$\frac{\log_2 v}{B}$ for $B$ bits payload
\end{itemize}

\section{Methods}

\subsection{Toy model}\label{sec:methods-toy-model}
We assume a two leaky integrators with linear feedforward coupling 
\begin{align}
\tau \frac{d}{dt} x_0(t) =& -x_0(t) + c \\
\tau \frac{d}{dt} x_1(t) =& -x_1(t) + x_0(t) \;,
\end{align}

with exact solutions
\begin{align}
x_0(t) =& c \left( 1 - e^{-\frac{t}{\tau}} \right) \\
x_1(t) =& c \left( 1 - e^{-\frac{t}{\tau}} \right) - t \frac{c}{\tau} e^{-\frac{t}{\tau}} \;.
\end{align}

\subsection{Numerical solvers}\label{sec:methods-solvers}

RK1 (forward Euler)

\begin{align}
k_0 =& f(t, x(t)) \\
x(t+h) =& x(t) + h k_0
\end{align}

\noindent RK2 (Ralston) \citep{ralston1962runge}

\begin{align}
k_0 =& f(t, x(t)) \\
k_1 =& f(t + \frac{2}{3}h, x(t) + \frac{2}{3}h k_0) \\
x(t+h) =& x(t) + h \left( \frac{1}{4} k_0 + \frac{3}{4} k_1 \right)
\end{align}

\noindent RK3 (Ralston) \citep{ralston1962runge}

\begin{align}
k_0 =& f(t, x(t)) \\
k_1 =& f(t + \frac{1}{2}h, x(t) + \frac{1}{2}h k_0) \\
k_2 =& f(t + \frac{3}{4}h, x(t) + \frac{3}{4}h k_1) \\
x(t+h) =& x(t) + h \left( \frac{2}{9} k_0 + \frac{1}{3} k_1 + \frac{4}{9} k_2 \right)
\end{align}

\subsection{Number of function evaluations}\label{sec:methods-nfevs}

\begin{align}
\notag
n_f(\epsilon, p) =& s \frac{T}{h(\epsilon, p)} \\
\sim& p T \epsilon^{-\frac{1}{p}} \;,
\end{align}
Technically the first term scales proportional to the number of stages $s$, not solver order $p$.
However $s$ is almost proportional to $p$ \citep{hairer1993solving}.

\subsection{Optimal order for given error tolerance}\label{sec:methods-optimal-order}
\begin{align}
0 = \frac{\partial}{\partial p} \left( p T \epsilon^{-\frac{1}{p}} \right) =& T \epsilon^{-\frac{1}{p}} + p T \epsilon^{-\frac{1}{p}} \log(\epsilon) \frac{1}{p^2} \\
0 =& 1 + \log(\epsilon) \frac{1}{p} \\
\Rightarrow p =& -\log(\epsilon)
\end{align}

\subsection{Optimal step size for finite precision}\label{sec:methods:optimal-step-size}

\begin{align}
0 = \frac{\partial}{\partial h} \left( aTh^p + \frac{bT}{h} \right) =& aT p h^{p-1} - \frac{bT}{h^2} \\
h^{p+1} =& \frac{b}{ap} \\
\Rightarrow h =& \left(\frac{c}{p}\right)^{\frac{1}{p+1}}
\end{align}

\subsection{Larger optimal step size for higher-order method}\label{sec:methods-larger-optimal-step-size}

\begin{align}
\frac{\partial}{\partial p} \left( \frac{c}{p} \right)^{\frac{1}{p+1}} =& \left( \frac{c}{p} \right)^{\frac{1}{p+1}} \left( -\frac{1}{(p+1)^2} \log \frac{c}{p} - \frac{1}{(p+1) p} \right) \\
=& -\left( \frac{c}{p} \right)^{\frac{1}{p+1}} \frac{1}{(p+1)^2 p} \left( p \log \frac{c}{p} + (p+1) \right) \\
\Rightarrow 0 >& p \log \frac{c}{p} + (p+1) \\
\log p - \frac{p+1}{p} >& \log c \\
p e^{-\frac{1}{p} - 1} > c
\end{align}

\begin{figure}[t]
\centering
\includegraphics[width=0.49\textwidth]{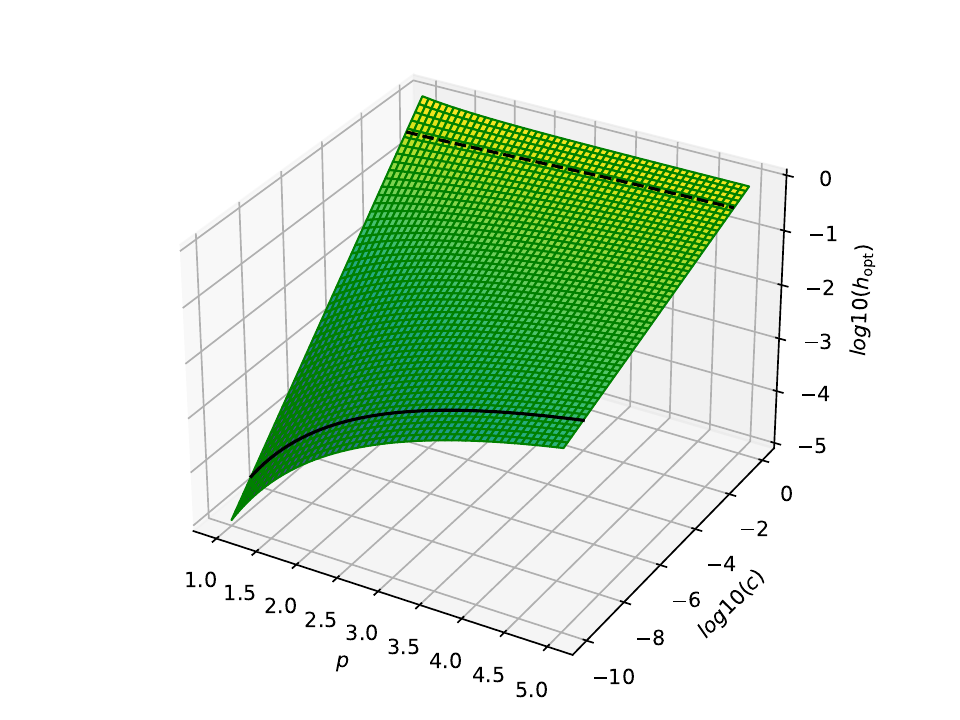}
\includegraphics[width=0.49\textwidth]{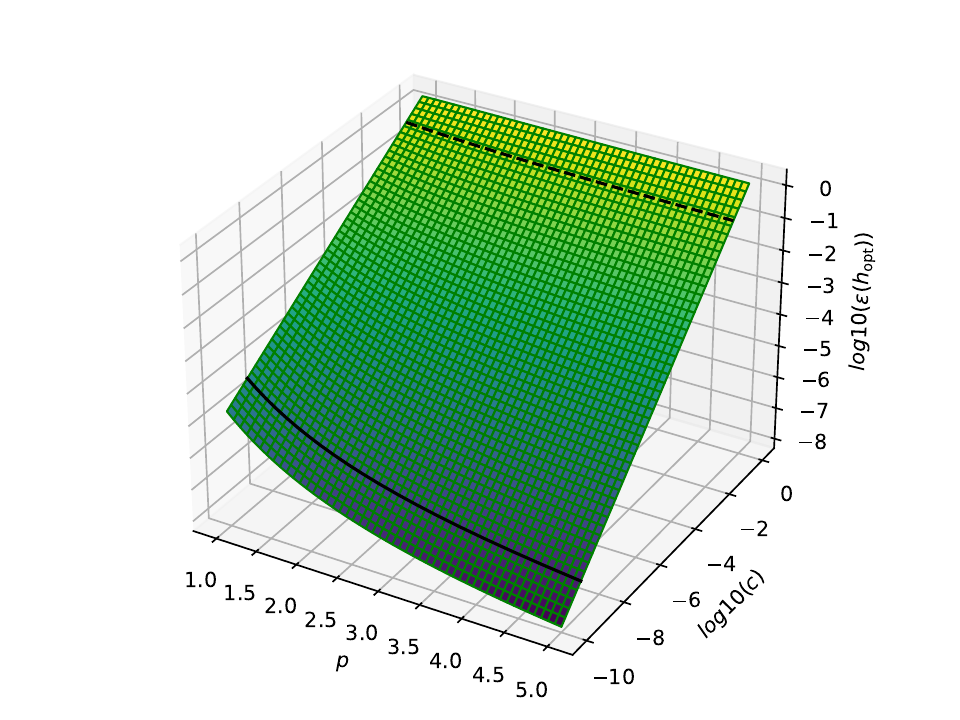}
\caption{{\bf Optimal step size (left) and error at optimal step size (right).}
Black solid line indicates $c=10^{-9}$, black dashed line indicates $c=0.135$.
Note that $p$ can only be meaningfully interpreted at integer values.
}\label{fig:methods-optimal-step-size-error}
\end{figure}

\subsection{Smaller error at optimal step size for higher-order method}\label{sec:methods-smaller-error}

\begin{align}
\frac{\partial}{\partial p}\Big|_{h_\text{min}^{(p)}} \left( aTh^p + \frac{bT}{h} \right) =& aTh^p \log h \Big|_{h_\text{min}^{(p)}} \\
=& aT \left(\frac{c}{p} \right)^{\frac{p}{p+1}} \frac{1}{p+1} \log \left( \frac{c}{p} \right) \\
\Rightarrow p >& c
\end{align}

\subsection{NeuroScale model}\label{sec:methods-neuroscale-model}
Each neuron is modeled as a leaky integrator with nonlinear activation function
\begin{align}
\tau_i \frac{d}{dt} x_i(t) = -x_i(t) + \sum_{j} w_{ij} \varphi(x_j(t)) + b_i
\end{align}
where $\varphi(x) = \text{ReLU}(x)$.

We simulate a network of $N=43$ neurons for a duration of $T=9.1$ time units, assigning $21$ neurons per core.

Each neuron is randomly assigned a parameter combination from
\begin{align*}
\{&\\
&[\tau=0.36, b=0.00825]\\
&[\tau=0.66, b=0.024]\\
&[\tau=0.96, b=0.01075]\\
\}&
\end{align*}

Weights drawn according to PyTorch's Kaiming initialization with fan-in equal to total number of neurons, scaled by $4.0$ and then shifted by $-0.2$, randomly downsampled to $25\%$ density and quantized to Q$4.12$ (for historical reasons stronger quantization than used in simulation).

\subsection{NeuroScale back-annotated simulations}\label{sec:methods-neuroscale-simulations}
Energy and delay for control elements are obtained by SPICE simulations and for combinational logic blocks from Cadence Genus.
We use AMC \citep{ataei2019amc} to generate memories and obtain energy and delay for read and write operations from SPICE simulations.

\end{document}